\documentstyle[11pt]{article}

\begin{document}

\newcommand{\bu}{{\bf u}} 
\newcommand{\bsig}{\mbox{\boldmath$\sigma$}}
\newcommand{\bmu}{\mbox{\boldmath$\mu$}}
\newcommand{\bchi}{\mbox{\boldmath$\chi$}}
\newcommand{\bpi}{\mbox{\boldmath$\chi$}_{p}}
\newcommand{\bxi}{\hat{\mbox{\boldmath$\xi$}}}
\newcommand{\bs}{\hat{\mbox{\boldmath$\chi$}}}
\newcommand{\bw}{\mbox{\boldmath$\omega$}}
\newcommand{\bbeta}{\mbox{\boldmath$\beta$}}
\newcommand{\bel}{\begin{equation}\label}
\newcommand{\ee}{\end{equation}} 
\newcommand{\I}{\int_{\Omega}}
\newcommand{\beq}{\begin{eqnarray}} 
\newcommand{\eeq}{\end{eqnarray}} 
\newcommand{\bc}{\begin{center}} 
\newcommand{\ec}{\end{center}} 
\newcommand{\bk}{{\bf {\hat k}}}
\newcommand{\cd}{\cdot}
\newcommand{\tm}{\times}
\newcommand{\ptl}{\partial}
\newcommand{\var}{\varepsilon}
\newcommand{\bwint}{\frac{1}{r^2}\int_{0}^{r}r'\,a(r',t)\,dr'}
\newcommand{\at}{\tilde{\alpha}}
\newcommand{\lt}{\tilde{\lambda}}
\newcommand{\mt}{\tilde{\mu}}
\newcommand{\aw}{\alpha^{(\omega)}}
\newcommand{\bchiw}{\mbox{\boldmath$\chi^{(\omega)}$}}

\begin{titlepage}
\date{}
\title{{\Large\bf Vorticity alignment results for the
three-dimensional Euler and Navier-Stokes equations}}
\author{\Large B. Galanti$^{\dag}$, J. D. Gibbon$^{*\dag}$ and 
M. Heritage$^{*}$} 
\maketitle 
\begin{center} 
$\dag$ Department of Chemical Physics,\\
Weizmann Institute of Science, 76100 Rehovot, Israel.\\
\vspace{0.1cm}
$*$ Department of Mathematics,\\
Imperial College of Science, Technology and Medicine,\\
London SW7 2BZ, UK.\\
\end{center} 
\thispagestyle{empty} 
\abstract\noindent
We address the problem in Navier-Stokes isotropic turbulence of why
the vorticity accumulates on thin sets such as quasi-one-dimensional
tubes and quasi-two-dimensional sheets.  Taking our motivation from
the work of Ashurst, Kerstein, Kerr and Gibson, who observed that the
vorticity vector $\mbox{\boldmath$\omega$}$ aligns with the
intermediate eigenvector of the strain matrix $S$, we study this
problem in the context of both the three-dimensional Euler and
Navier-Stokes equations using the variables $\alpha =
\hat{\mbox{\boldmath$\xi$}}\cdot S\hat{\mbox{\boldmath$\xi$}}$ and
$\mbox{\boldmath$\chi$} = \hat{\mbox{\boldmath$\xi$}}\times
S\hat{\mbox{\boldmath$\xi$}}$ where $\hat{\mbox{\boldmath$\xi$}} =
\mbox{\boldmath$\omega$}/\omega$. This introduces the dynamic angle
$\phi ({\bf x},t) = \arctan\left(\frac{\chi}{\alpha}\right)$, which
lies between $\mbox{\boldmath$\omega$}$ and
$S\mbox{\boldmath$\omega$}$.  For the Euler equations a closed set of
differential equations for $\alpha$ and $\mbox{\boldmath$\chi$}$ is
derived in terms of the Hessian matrix of the pressure $P =
\{p_{,ij}\}$. For the Navier-Stokes equations, the Burgers vortex and
shear layer solutions turn out to be the Lagrangian fixed point
solutions of the equivalent $(\alpha,\mbox{\boldmath$\chi$})$
equations with a corresponding angle $\phi = 0$.  Under certain
assumptions for more general flows it is shown that there is an
attracting fixed point of the $(\alpha,\bchi)$ equations which
corresponds to positive vortex stretching and for which the cosine of
the corresponding angle is close to unity. This indicates that near
alignment is an attracting state of the system and is consistent with
the formation of Burgers-like structures.
\begin{center}
email addresses: galanti@dvir.weizmann.ac.il; j.gibbon@ic.ac.uk
{\bf AMS:} 76F05, 76C05  
\end{center}

\end{titlepage}


\section{Introduction}

As early as 1938 G. I. Taylor \cite{GIT1} showed that in isotropic
Navier-Stokes turbulence vortex stretching has a major effect on
vorticity production and dissipation. One of the many interesting
features of high Reynolds number turbulent flows, illustrated
beautifully by modern flow visualization methods, is the fact that
vorticity is not evenly distributed throughout a flow but has a
tendency to accumulate on `thin sets'.  The morphology of these sets
is characterized by the predominance of quasi one-dimensional tubes or
filaments and quasi two-dimensional sheets although, being fractal in
nature \cite{Chorin1,Frischbk,beta}, many individual vortical
structures may be neither precisely one nor the other.  This
morphology makes it clear that the {\em direction} of vorticity as
well as its production plays a significant role in the
self-organization processes through which apparently intense regions
of vorticity appear to metamorphose into approximate thin coherent
structures. The challenge facing the theorist working with the
three-dimensional Navier-Stokes equations is to explain these
geometric accumulation processes and their consequences even though a
detailed description of their precise topology is beyond our reach at
present. The subtle fine spatial structure of these sets indicates
that an alternative approach is needed to the conventional one of
Navier-Stokes analysis where $L^{2}$-spatial averages are taken over
the whole domain when estimating the effect of the vortex stretching
term.

In 1951, Townsend \cite{Town1,GKB1} indicated that something like the
thin sets described above might be relevant in turbulent flows and in
the last two decades their occurrence has been a regular theme of the
literature
\cite{Sig1,HKM2,HKM3,Ld1,LC1,Neu1,Mjd1,AKKG,ACR,She1,HY2,VM1,Mjd2,RM,
Cd1,Kida1,Kida2,Saffbk,MKO,VM2,Sulem1,BMC}.  They would also seem to
be the visual manifestation of the vorticity alignment phenomenon
first reported by Ashurst {\em et al} \cite{AKKG} who observed that in
driven simulated isotropic Navier-Stokes turbulence with a Taylor
microscale Reynolds number of 83 the vorticity vector $\bw$ has a
tendency to align with the intermediate eigenvector of the rate of
strain matrix
\bel{Sdef}
S_{ij} = \frac{1}{2}\left(u_{i,j} + u_{j,i}\right). 
\ee
This observation was based on a study of the probablility density
function (PDF) of the cosine of the angle between their
directions. This important alignment process has also been confirmed
in other numerical simulations \cite{ACR,She1,VM1,RM,VM2,BMC} as well
as in turbulent grid flow experiments conducted by Tsinober {\em et
al} \cite{ATS1,ATS2,ATS3,ATS4} and it has been suggested that it is a
kinematic process \cite{Jim,DT1}. In fact preferential vorticity
alignment appears to be a universal feature of smaller scale
structures in such flows even though larger scale features may vary
from flow to flow. A common feature seems to be that vorticity is
concentrated in tubes of width intermediate between the Taylor and
Kolmogorov microscales \cite{AKKG,She1,VM1,RM,BMC} with viscous
dissipation occurring in annular regions around the tubes which
themselves form from the roll-up of vortex sheets \cite{VM2}.

In isolation from each other and in an ideal sense, vortex tubes and
sheets can be thought of as Burgers vortices and Burgers (stretched)
shear layers respectively for which there are known exact solutions of
the Navier-Stokes equations in certain special cases
\cite{MKO,Mjd1,Saffbk,Burg1}. Of course the Burgers solutions are
idealized because their strain fields are not linked back to the local
vorticity field as would be the case in a real flow.  Nevertheless
they have the great merit of providing us with exact solutions which
allow an interpretation of the processes involved. A Burgers vortex
tube ideally has one positive and two negative eigenvalues of the
strain matrix (axial strain) while a Burgers shear layer solution has
two positive and one negative eigenvalue (biaxial strain). Ashurst
{\em et al} \cite{AKKG} reported that their eigenvalues occurred in
the ratio $3:1:-4$~~but, as Moffatt, Kida and Ohkitani have pointed
out \cite{MKO}, vortex tubes can still survive in regions of biaxial
strain provided they are strong enough. As best illustrated by the
graphics of Vincent and Meneguzzi \cite{VM2}, vortex sheets deform
neighbouring sheets, curling up like potato crisps (chips) most
strongly when rolling up into tubes. It is significant that they found
that the tendency towards alignment between $\bw$ and the intermediate
eigenvector of $S$ occurred before the roll-up of sheets into tubes.
Burgers vortices have been produced in the laboratory by Andreotti
{\em et al} \cite{Cd2} from two experimental set-ups which allow them
to study the phenomenon of vortex stretching in detail.  While the
Burgers solutions are extremely valuable to the theorist, it is
necessary to formulate a theory which demonstrates why a flow should
evolve to such states in the first place.

For the three-dimensional Euler equations the situation is somewhat
different and complicated by the suspicion of a finite time
singularity in the vorticity field (see Kerr \cite{Kerr2,Kerr3}).
Constantin, Fefferman and Majda \cite{CFM,Constantinreview}, however,
have recently shown that singularities in solutions of the
three-dimensional incompressible Euler equations can be ruled out if
the velocity is finite and the direction of vorticity is smooth but
they cannot be ruled out if the direction of vorticity is not
smooth. For instance, blow-up cannot be ruled out if two intense
vortex tubes collide at a nontrivial angle.  In fact, Kerr's
computations predicting such a singularity used anti-parallel vortex
tubes as initial data \cite{Kerr2,Kerr3} (see also
\cite{BP1,Cantwell}). The result of Constantin, Fefferman and Majda
\cite{CFM} is related to that of Beale, Kato and Majda \cite{BKM}
which has provided the main criterion for understanding the growth of
vorticity in Euler flows for more than a decade.

We begin by setting up the notation and then proceed to give some
preliminary definitions.  Let us consider the incompressible 
three-dimensional Euler equations in vorticity form
\bel{eul1}
\frac{D\bw}{Dt}=\bsig , 
\ee
and the Navier-Stokes equations 
\bel{eulNS}
\frac{D\bw}{Dt}=\bsig + \nu \Delta \bw,
\ee
with a divergence free velocity field $\mbox{div}\,\bu = 0$.  The 
material derivative is defined in the usual manner as 
\bel{Ddef}
\frac{D~}{Dt} = \frac{\partial~}{\partial t}+\bu\cdot\nabla .
\ee
The vortex stretching vector $\bsig$ appearing in (\ref{eul1}) and
(\ref{eulNS}) is given by
\bel{eul2}
\sigma_{i} = \omega_{j}u_{i,j} = S_{ij}\omega_{j}.
\ee
$\bsig$ is written as $\bsig = S\bw$ on the right hand side of
(\ref{eul2}) because $\bw$ sees only the symmetric part of
$u_{i,j}$. The approach taken in this paper is to consider the
dynamics of the local angle $\phi ({\bf x},t)$ which lies between the
vorticity vector $\bw$ and $S\bw$ at the point ${\bf x}$.  Not having
independent evolution equations for the eigenvectors of $S$ restricts
our ability to interpret the geometry of the problem. The significance
of $\phi$ in this context is that when it takes the value(s) $\phi
=0~(\pi)$ this means that $\bw$ has aligned (anti-aligned) with an
eigenvector of $S$, although which one we cannot say. The so-called
`stretching rate' $\alpha$ is an interesting quantity to consider
($\bxi = \bw/|\bw |$)
\bel{alphadef}
\alpha ({\bf x},t) = \frac{\bw\cd\bsig}{\bw\cd\bw} = 
\frac{\bw\cd S\bw}{\bw\cd\bw} = \bxi\cd S\bxi,
\ee
which has been related to the vorticity through an elegant Biot-Savart
integral \cite{Constantinreview,CFM}. For the three-dimensional 
Euler equations, the scalar vorticity $\omega = |\bw|$ simply obeys
\bel{const}
\frac{D\omega}{Dt}=\alpha \omega .
\ee
The scalar $\alpha$ is obviously an estimate for an eigenvalue of the
strain matrix $S$ and lies within its spectrum. By defining the vector 
\bel{chidefintro}
\bchi ({\bf x},t) 
= \frac{\bw\times S\bw}{\bw\cdot\bw} = \bxi\tm S\bxi
\ee
the angle $\phi ({\bf x},t)$ between $\bw$ and $S\bw$ can 
naturally be introduced as $(\bchi\cdot\bchi = \chi^{2})$
\bel{ang1}
\tan\phi ({\bf x},t) = \frac{|\bw\times S\bw|}{\bw\cdot S\bw}
=\frac{\chi}{\alpha}
\ee
and its dynamics related to those of $\alpha$ and $\bchi$.  Gibbon and
Heritage \cite{GHang1} have recently discussed this angle but only in
the context of the volume average over the whole flow.  The stretching
rate $\alpha$ takes account only of the symmetric part $S$ of the
deformation matrix $u_{i,j}$ but the antisymmetric part clearly
contributes to $\bchi$.  For a fluid element at a point ${\bf x}$
certain specific values of the angle $\phi$ can be interpreted as the
following:

\begin{enumerate}

\item When $\phi = 0$ the vectors $\bw$ and $S\bw$ are parallel and
only stretching occurs. 

\item When $\phi = \pi/2$ the vectors $\bw$ and $S\bw$ are orthogonal
and, since $S\bw = D\bw/Dt$, this means that in this case {\em $\bw$
rotates but does not stretch}. This effect will nevertheless distend
and misalign vortex lines.

\item The case $\phi = \pi$ represents anti-alignment. In this case
$\alpha < 0$, so vorticity collapses rapidly to small values. 

\item When $0 < \phi < \pi/2$ both stretching and rotation occur
simultaneously at each point on a vortex line.

\end{enumerate}
In order to find the most sensitive relationship between $\alpha$ and
$\bchi$ we take advantage in \S2 of a result of Ohkitani
\cite{Ohk1,Ohk2} for the incompressible three-dimensional Euler
equations
\bel{sigderiv}
\frac{D\bsig}{Dt} = -P\bw ,
\ee
where $P =\{p_{,ij}\}$ is the Hessian matrix of the pressure.  This
result\footnote{Note added in proof: In a private communication,
P. Constantin has informed us that the result in equation (10) has
been known privately in unpublished form since 1983.} comes about
from a more general well known property of the Euler equations, namely
that if $\bu$ and $\bw$ are velocity and vorticity solutions of the
Euler equations then for any arbitrary vector ${\bf A}$
\bel{Aequn}
\frac{D}{Dt}\left(\omega_{j}\frac{\partial A_{i}}{\partial x_{j}}\right) 
= \sigma_{j}\frac{\partial A_{i}}{\partial x_{j}} 
- \sigma_{k}\frac{\partial A_{i}}{\partial x_{k}}
+ \omega_{j}\frac{\partial}{\partial x_{j}}\frac{DA_{i}}{Dt}
\ee
from which we conclude that
\bel{Aequn2}
\frac{D}{Dt}(\bw\cdot\nabla {\bf A}) 
= \bw\cdot\nabla \left(\frac{D{\bf A}}{Dt}\right).
\ee
Choosing ${\bf A} = \bu$ with a direct use of the velocity form of the
Euler equations on the right hand side of equation (\ref{Aequn2})
immediately gives equation (\ref{sigderiv}).  While the pressure
Hessian in (\ref{sigderiv}) is a fully nonlinear term, ($\mbox{tr}P$
is related to $u_{i,j}$ by $\Delta p = - u_{i,j}u_{j,i}$),
nevertheless the cancellation of two terms inherent in the derivation
of equation (\ref{sigderiv}), each of which are
$\sim\bw|\nabla\bu|^{2}$, directly removes all the terms not
explicitly dependent on $P$.  The approach of this paper therefore
runs counter to the conventional one where the pressure is removed by
projection but the advantage gained by the fortuitous cancellation of
nonlinear terms comes at the price of having to deal with the
$P$-matrix. Previous studies on depletion of nonlinearity have been
made by Constantin and Fefferman \cite{CF1} for the three-dimensional
Navier-Stokes equations and Constantin, Fefferman and Majda \cite{CFM}
for the three-dimensional Euler equations who have considered the
angle between the vorticity vectors $\bw ({\bf x})$ and $\bw ({\bf
y})$ at two points ${\bf x}$ and ${\bf y}$ in the flow.

The main results of the paper can be summarized as follows. It is
shown in \S2 that at points in the flow for which $\bw \neq
0$, the material derivatives of $\alpha$ and $\bchi$ for the Euler
equations obey
\bel{alphaev}
\frac{D\alpha}{Dt} = \chi^{2} - \alpha^{2} - \alpha_{p},
\ee 
\bel{chi0}
\frac{D\bchi}{Dt} = -2\alpha \bchi - \bpi,
\ee
where $\alpha_{p}$ and $\bchi_{p}$ are related to the Hessian matrix
$P$ which is a non-local quantity in the flow
\cite{Ohk2,VF1,WPCheng}. For the Navier-Stokes equations (see \S3) the
addition of viscosity produces differential equations of the type
\bel{extra1}
\frac{D\alpha}{Dt} = \chi^{2} - \alpha^{2} + \nu \Delta \alpha 
+2\nu \alpha |\nabla\bxi |^{2}+\lambda ,
\ee 
\bel{extra2}
\frac{D\bchi}{Dt} = -2\alpha \bchi + \nu\Delta\bchi 
+2\nu\bchi|\nabla\bxi |^{2} + \bmu ,
\ee
where the terms in $|\nabla\bxi |^{2}$ express the misalignment of
vortex lines in the differential geometric sense \cite{CPS,GPS}. The
terms $\lambda$ and $\bmu$ contain information about the flow and are
not generally constant.  In \S4, however, the exact solutions of the
Navier-Stokes equations that represent Burgers vortex tubes and shear
layers are discussed and it is discovered there that these correspond
to `Lagrangian fixed point' solutions of (\ref{extra1}) and
(\ref{extra2}) in the sense that $D\alpha/Dt = 0$ and $D\bchi/Dt = 0$,
$\lambda$ and $\alpha$ are positive real constants and $\bmu = 0$ and
$\bchi = 0$. Hence for both these thin structures $\lambda$ and $\bmu$
simplify greatly and the corresponding angle $\phi$ is exactly zero.
Using this as motivation, it is assumed in \S5 that if the flow is
regular then all the variables must come to some equilibrium in a
connected region of high vorticity.  From this we want to see if the
equilibrium values of $\alpha$ and $\bchi$ correspond to a {\em small}
angular orientation $\phi_{0}$. In order to do this it is assumed that
$\lambda$ and $\bmu$ are constant and then it is shown that equations
(\ref{extra1}) and (\ref{extra2}) have two fixed point solutions
$(\alpha_{0},\bchi_{0})$ in the Lagrangian sense, one of which is
repelling ($\alpha_{0} < 0$) while the other ($\alpha_{0} > 0$) is
attracting.  The stability of the positive root for $\alpha$ shows
that the system favours vortex stretching, in agreement with
G. I. Taylor's conclusions \cite{GIT1}. For $\lambda > 0$, the stable
solution has a corresponding small attracting angle $\phi_{0}$ which
is insensitive to the relative ratio of $\lambda$ and $|\bmu|$, making
the natural orientation of the vectors close to true alignment.  These
results are consistent with near Burgers-like structures forming in
the flow.  It remains to be proved, however, that the Burgers
solutions are either unique or belong a more general unique class of
solutions that correspond to thin sets which are true Lagrangian fixed
points with small values of $\phi$. As we point out in \S2 neither
Hill's spherical vortex (an exact solution of the Euler equations) nor
ABC flow belong to the class we are considering as neither are
Lagrangian fixed points of their respective ($\alpha,\bchi$)
equations.

\section{Stretching equations for the three-dimensional Euler equations}

The scalar $\alpha$ and the vector $\bchi$ were defined in equations
(\ref{alphadef}) and (\ref{chidefintro}) by forming the dot and cross
products respectively of $\bw$ with $ S\bw$. We repeat their
definitions:
\bel{alphachidef}
\alpha = \frac{\bw\cd S\bw}{\bw\cd\bw} = \bxi\cd S\bxi,\hspace{1.5cm}
\bchi = \frac{\bw\times S\bw}{\bw\cd\bw} = \bxi\times S\bxi ,
\ee
but exclude points where $\bw =0$. The unit vector $\bs({\bf x},t)$, 
defined by
\bel{etadef}
\bs = \frac{\bchi}{\chi},
\ee 
is orthogonal to the unit vector $\bxi$; for example, if $\bxi$ points
along a vortex tube then $\bs$ lies in the plane orthogonal to the
tube.  Let us also define the scalar quantity $\alpha_{p}$ and the
vector $\bpi$
\bel{evSP}
\alpha_{p}({\bf x},t) =\frac{\bw\cd P\bw}{\bw\cd \bw}
\hspace{2cm}
\bpi ({\bf x},t) = \frac{\bw\times P\bw}{\bw\cd \bw}.
\ee
$\alpha_{p}$ is an estimate for an eigenvalue of
the Hessian of the pressure $P=\{p_{,ij}\}$ in the same way that
$\alpha$ is for the strain matrix $S$. When $\bw$ aligns with an
eigenvector of $P$ then $\alpha_{p}$ becomes an exact eigenvalue of
$P$ whereas $\bpi$ becomes zero in this case.

In \S1 the definitions (\ref{alphadef}) and (\ref{chidefintro}) were
used to define the angle $\phi$ between $\bw$ and $\bsig =S\bw$,
namely $\tan\phi = \chi/\alpha$, as a way of characterizing the
alignment or misalignment between them.  In terms of the vector $\bs$
this angle obeys ($\bsig = S\bw$)
\bel{phi1}
\bs\tan\phi = \frac{\bw\times\bsig}{\bw\cdot\bsig}
= \frac{\bw\times S\bw}{\bw\cd S\bw},
\ee
with $\phi$ varying in the range 
$0 \leq \phi\leq 2\pi$.  One consequence of equation (\ref{phi1}) 
is that
\bel{cos}
\cos^{2}\phi = \frac{(\bsig\cd \bw)^{2}}{|\bsig|^2\,|\bw|^2} 
= \frac{\omega^{2}\alpha^{2}}{\sigma^{2}}.
\ee
To obtain relations between these quantities we firstly note that
\bel{conv1}
\frac{D (\bw\cd\bsig )}{Dt} = \sigma^{2} -\bw\cdot P\bw 
= \omega^{2}\left( \alpha^{2}\sec^{2}\phi -\alpha_{p}
\right).
\ee
whereas in the cross product, the terms in $\bsig$ vanish leaving 
only
\bel{conv2}
\frac{D (\bw\tm\bsig )}{Dt} =  -\bw\tm P\bw .
\ee
Equation (\ref{conv2}) shows that in addition to the cancellation of
nonlinear terms in (\ref{sigderiv}), more nonlinearity has been
lost through the cross product. The material derivatives of $\alpha$
and $\bchi$ can now easily be obtained through equations
(\ref{alphadef}), (\ref{cos}), (\ref{conv1}) and (\ref{conv2}),
\bel{alphaev1}
\frac{D\alpha}{Dt} = \chi^{2} - \alpha^{2} - \alpha_{p},
\ee 
\bel{chiidef}
\frac{D\bchi}{Dt} = -2\alpha \bchi - \bpi.
\ee
Subject to solutions existing, the prominent feature of
(\ref{alphaev1}) and (\ref{chiidef}) is that they are independent
explicitly of $\bw$ and $\phi$ while being driven by $\alpha_{p}({\bf
x},t)$ and $\bpi ({\bf x},t)$.  Hence they are a set of non-autonomous
ODEs, in the Lagrangian picture of fluid mechanics, operating in a
4-dimensional phase space $(\alpha, \chi_{i})$ for $i=1,2,3$. The
three equations in the $\chi_{i}$ can be reduced to one in the scalar
$\chi = \sqrt{\bchi\cdot\bchi}$ to give the pair of equations
\bel{alphaev2}
\frac{D\alpha}{Dt} = \chi^{2} - \alpha^{2} - \alpha_{p},
\ee 
\bel{chi1}
\frac{D\chi}{Dt} = -2\alpha \chi - \tilde{\chi}_{p},
\ee
where
\bel{chipdef}
\tilde{\chi}_{p} = \bs\cdot\bpi .
\ee
On the $\alpha$-axis where $\phi =0$ or $\pi$, equation (\ref{conv2})
show that $\bw\times P\bw = 0$.  Hence $\bchi_{p} = 0$ and so
$D\bchi/Dt = 0$ on this axis. Consequently equation (\ref{chi1})
operates only in the the upper half-plane. The lower half-plane is
simply a reflection of the upper in the $\alpha$-axis. While $\alpha$
is the stretching rate for the scalar vorticity $\omega$, equation
(\ref{chi1}) shows that $2\alpha$ is the contraction rate for $\chi$
with an additional pressure term. The two scalars $\omega$ and $\chi$
are also related by
\bel{omegachi}
\frac{D^{2}\omega}{Dt^{2}} 
= \left(\chi^{2} - \alpha_{p}\right)\omega .
\ee
The pair of equations for $(\alpha,\chi )$ given above can be merged
into one by defining the complex stretching rate 
\bel{zetadef}\
\zeta = \alpha + i\chi
\ee
and (\ref{alphaev2}) and (\ref{chi1}) reduce to
\bel{zeta1}
\frac{D\zeta}{Dt} +\zeta^{2} + \zeta_{p} = 0.
\ee
$\zeta_{p}$ is simply the complex combination $\zeta_{p} = \alpha_{p}
+i\tilde{\chi}_{p}$. Because the $\zeta$-plane is only the upper half
of the complex plane, $\phi$ is valid only in the range $0\leq
\phi\leq\pi$.  Two related results, which are consequences of
equations (\ref{alphaev2}) and (\ref{chi1}), are
\bel{zeta2}
\frac{D(|\zeta|^2)}{Dt} = -2\alpha |\zeta|^{2} 
-2\Re\left\{\zeta\zeta_{p}^{\ast}\right\},
\ee
and
\bel{tan2}
\frac{D\,(\tan\phi)}{Dt} = -\alpha\tan^{3}\phi -\left(\alpha 
-\frac{\alpha_{p}}{\alpha}\right)\tan\phi 
- \frac{\tilde{\chi}_{p}}{\alpha} .
\ee
Because of the negative cubic term, there is always the tendency in
the $\alpha > 0$ quarter-plane for the angle to decrease unless the
pressure terms force it to behave to the contrary.  Equations
(\ref{alphaev2}) and (\ref{chi1}), however, are non-autonomous and,
for Euler flows, there can be no question of $\alpha_{p}$ and
$\tilde{\chi}_{p}$ remaining constant as the flow
develops. Nevertheless certain tendencies can be discussed:

\begin{enumerate}

\item In the left hand quarter-plane equation where $\alpha < 0$
(\ref{const}) shows that vorticity must be very small here. Initially
if one starts in the region $\alpha < 0$ then equation (\ref{chi1})
shows that $\chi$ can undergo rapid growth with
$\phi\rightarrow\pi/2$.  This is also reflected in equation
(\ref{zeta2}) and equation (\ref{tan2}) where the terms $-2\alpha
|\zeta|^2$ and $-\alpha\tan^{3}\phi$ force rapid growth in $|\zeta|$
and $\tan\phi$ respectively. If $\chi$ increases in this way then
(\ref{alphaev2}) shows that there is also the tendency for $\alpha$ to
increase also. Consequently there is a natural tendency for orbits to
pass from the left to the right hand half plane unless the pressure
acts to prevent it.

\item In the $\alpha > 0$ quarter-plane the opposite process
occurs. Here, if $\alpha_{p} < \alpha^2$ and $\tilde{\chi}_{p}> 0$ for
a sufficiently long time then $\phi \rightarrow 0$, which is the state
of exact alignment.  If $\alpha_{p} > \alpha^2$ then how far $\phi $
decreases depends upon specific values of $\alpha_{p}$ and
$\tilde{\chi}_{p}$.

\end{enumerate}
We conclude that the system has a natural tendency to reject negative
values of $\alpha$ unless forced by the pressure Hessian to behave
otherwise.  This means that the system tends to prefer regions where
stretching is positive.  It is noteworthy that in rejecting the
$\alpha < 0$ region an orbit must pass through the $\chi$-axis into
the region of $\alpha > 0$.  This process can be reversed if the
pressure behaves in a contrary manner but this behaviour can only be
properly elucidated from data and this analysis is in progress.

What of exact solutions of the Euler equations? As we shall see in
\S4, the Burgers vortex solution of the Euler equations is a
Langrangian fixed point of (\ref{alphaev2}) and (\ref{chi1}) with
$\alpha = \mbox{const},~\chi = 0$ with a corresponding angle $\phi
=0$.  Hill's spherical vortex, on the other hand, also has $\phi =0$
but this does {\em not} correspond to a Lagrangian fixed point as
fluid packets do not have values of $\alpha$ for which $D\alpha/Dt =
0$, based on the fact that $\bu\cdot\nabla\alpha \neq 0$.  To see this
consider a vorticity field $\bw = (\omega_{r},
\omega_{\theta},\omega_{z})$ where $\omega_{r} =\omega_{z}=0$ and
$\omega_{\theta} = Ar$ inside the sphere and zero outside it
\cite{GKB2}. Then $\alpha = Az/5$, $\bchi =0$, $\alpha_{p} =
A^{2}\left[ 4r^{2}-\frac{10a^{2}}{3}\right]/50$ and $\bchi_{p}=0$.
The solution for ABC flow for which $\bu =\bw$ with $u_{1} =\omega_{1}
= \sin z + \cos y,~~u_{2} = \omega_{2} = \sin x + \cos z,~~u_{3} =
\omega_{3} = \sin y + \cos x$, does not have constant values for
$\alpha$ or $\bchi$. This means that neither $\phi$ is constant nor
are the material derivatives of $\alpha$ and $\bchi$ zero.

\section{The stretching variables and the Navier-Stokes equations}

For the Navier-Stokes equations, the task is to see how the addition
of viscosity changes the two equations for $\alpha$ and $\bchi$ in
(\ref{alphaev1}) and (\ref{chiidef}).  In vorticity form the
Navier-Stokes equations are
\bel{Dw2}
\frac{D\omega_{j}}{Dt} = S_{jk}\omega_{k} + \nu \Delta\omega_{j},
\ee
and the equivalent of (\ref{const}) for the scalar vorticity $\omega =
|\bw|$ is
\bel{Dw1}
\frac{D\omega}{Dt} = \alpha \omega + \nu \bxi\cd\Delta\bw .
\ee
>From the definition $\bxi = \bw/\omega$
\bel{Dxi/Dt}
\frac{D\xi_{j}}{Dt} = 
S_{jk}\xi_{k}-\alpha\xi_{j}+\frac{\nu}{\omega}\Delta 
\omega_{j}-\frac{\nu}{\omega}(\bxi\cdot\Delta\bw )\xi_{j}.
\ee
It is convenient to turn derivatives of $\bw$ into derivatives of
$\bxi$.  To achieve this we use the fact that $\xi_{j}^{2}=1$ leading
to $\xi_{j}\Delta\xi_{j} + |\nabla\xi_{j}|^{2} = 0$.  From the
relations $\omega = \omega_{j}\xi_{j}$ and $\omega_{j} =
\omega\xi_{j}$ we have
\bel{Dw4}
\Delta\omega = \omega |\nabla\bxi|^{2}+\bxi\cd\Delta\bw,
\ee
and
\bel{Dw5}
\frac{\Delta\omega_{j}}{\omega} = \Delta \xi_{j} 
+2\nabla (\ln \omega )\cd\nabla\xi_{j} 
+ \xi_{j}\frac{\Delta\omega}{\omega}.
\ee
Dividing (\ref{Dw4}) by $\omega$, multiplying by $\xi_{j}$ and then
adding to (\ref{Dw5}) enables us to rewrite (\ref{Dxi/Dt}) as   
\bel{Dxi/Dt,2}
\frac{D\xi_{j}}{Dt}=S_{jk}\xi_{k}-\alpha\xi_{j}+\nu\Delta\xi_{j}
+\nu|\nabla \bxi|^{2}\xi_{j}+\nu \beta_{j}
\ee
where the vector $\bbeta$ has components
\bel{djdef}
\beta_{j} = \frac{\ptl}{\ptl x_{k}}(\ln |\bw|^{2})
\frac{\ptl \xi_{j}}{\ptl x_{k}}.
\ee
$|\nabla\bxi|$ is the `misalignment' of the vector $\bxi$ in the
Frenet-Serret frame (see references \cite{CPS,GPS} and the calculation
in Appendix 1).  Equation\footnote{This expression is not valid at
stagnation points where $\bw = 0$ neither is $\bxi$ defined at these.}
(\ref{djdef}) is the key to the calculations in the rest of this
section because, apart from the logarithmic derivative in $\omega$, it
contains only $\bxi$ and its derivatives. The strain matrix evolves
according to
\bel{strain,deriv2}
\frac{DS_{ij}}{Dt}=-S_{ik}S_{kj}-\frac{\omega_{i}\omega_{j}}{4}+
\frac{|\bw|^{2}\delta_{ij}}{4}-P_{ij}+\nu\Delta S_{ij},
\ee
and so,
\begin{eqnarray}\label{deriv,sig}
\frac{D}{Dt}(S_{ij}\xi_{j}) & = & - 
P_{ij}\xi_{j}-\alpha S_{ij}\xi_{j}\nonumber\\
&+& \nu \left\{ (\Delta S_{ij})\xi_{j}+ S_{ij}\Delta\xi_{j}
+ |\nabla \bxi|^{2}S_{ij}\xi_{j} +  S_{ij}\beta_{j}\right\}.
\end{eqnarray}
This, together with (\ref{Dxi/Dt,2}), can be combined to give,
\begin{eqnarray}\label{Dalpha/Dt}
\frac{D}{Dt}(\xi_{i}S_{ij}\xi_{j}) &=& \chi^{2}-\alpha^{2}+ 
2\nu|\nabla \bxi|^{2}\alpha \nonumber\\  
&+&\nu \left\{\xi_{i}(\Delta S_{ij})\xi_{j}
+ \xi_{i}S_{ij}\Delta\xi_{j} 
+ \Delta\xi_{i}S_{ij}\xi_{j}\right\}\nonumber\\  
&+& \nu\left\{\xi_{i}S_{ij}\beta_{j} +\beta_{i} S_{ij}\xi_{j}\right\}
- \xi_{i}P_{ij}\xi_{j}.
\end{eqnarray}
Out of the Laplacian terms in (\ref{Dalpha/Dt}) it is desirable to 
form $\Delta \alpha$ 
\bel{DalphaDt}
\frac{D\alpha}{Dt} = \chi^{2}-\alpha^{2}+
\nu \Delta \alpha + 2\nu |\nabla \bxi|^{2}\alpha + \sum A_{jj},
\ee
where 
\bel{Adef}
A_{jl} = \nu\left(\xi_{i}S_{ij}\beta_{l} 
+ \beta_{i}S_{ij} \xi_{l}\right) -2\nu T_{jl} - \xi_{j}P_{ij}\xi_{l},
\ee
and where $T_{jl}$ is defined as 
\bel{Tdef}
T_{jl} = \frac{\ptl \xi_{i}}{\ptl x_{k}}
\frac{\ptl S_{ij}}{\ptl x_{k}}\xi_{l}
+\frac{\ptl \xi_{i}}{\ptl x_{k}}S_{ij}\frac{\ptl \xi_{l}}{\ptl x_{k}} 
+ \xi_{i}\frac{\ptl S_{ij}}{\ptl x_{k}}\frac{\ptl \xi_{l}}{\ptl x_{k}} ,
\ee
with $\beta_{i}$ is defined in (\ref{djdef}). We have isolated in
(\ref{DalphaDt}) as many specific terms in $\alpha$ as possible and
in forming the Laplacian term $\Delta\alpha$ we have separated the
highest derivatives of $S$ from the rest which lie in $T_{jl}$.  For
$\bchi$ a similar result holds which is
\bel{Dchi1Dt}
\frac{D\chi_{i}}{Dt}  =  - 2\alpha \chi_{i} + \nu\Delta \chi_{i}
+ 2\nu |\nabla\bxi|^{2}\chi_{i} + \varepsilon_{ijk}A_{kj}.
\ee
In summary, if we define
\bel{mulambdadef}
\mu_{i} = \varepsilon_{ijk}A_{kj},\hspace{1cm}\mbox{and}
\hspace{1cm}\lambda = \sum A_{jj},
\ee
then equations (\ref{DalphaDt}) and (\ref{Dchi1Dt}) become 
\bel{DalphaDtlambda}
\frac{D\alpha}{Dt} = \chi^{2}-\alpha^{2}+
\nu \Delta \alpha + 2\nu |\nabla \bxi|^{2}\alpha + \lambda,
\ee
and 
\bel{Dchi2Dt}
\frac{D\bchi}{Dt}  =  - 2\alpha \bchi + \nu\Delta \bchi
+ 2\nu |\nabla\bxi|^{2}\bchi + \bmu .
\ee
As in the Euler equations in \S2 this can be reduced to one 
equation in the scalar $\chi$ 
\bel{Dchi3Dt}
\frac{D\chi}{Dt} = -2 \chi\alpha + \nu \Delta\chi + 
\nu \left(2|\nabla\bxi|^{2} - |\nabla\bs|^{2}\right)\chi
+ \tilde{\mu} ,
\ee
where 
\bel{tildemudef}
\tilde{\mu} = \bs\cdot\bmu .
\ee

\section{Burgers vortex and shear layer solutions}

We now postpone our discussion of the $(\alpha,\bchi)$ equations for
the Navier-Stokes case until \S5 and return to the Burgers vortex and
stretched shear layer solutions.  It is pertinent to ask whether
equivalent Langrangian fixed point solutions exist for both the Euler
and Navier-Stokes equations?  Despite the caveats made in \S1, the
simplicity of the Burgers solutions makes them candidates for this.
In the following two subsections the basic formulae for the
axisymmetric Burgers vortex\footnote{Note added in proof: See also the
paper by Andreotti B. (1997) ``Studying Burgers' models to investigate
the physical meaning of the alignments observed in turbulence,''
Phys. Fluids A {\bf 9}, 735-42.}  (see \cite{MKO,Mjd1,Saffbk,Burg1})
and the Burgers shear layer (see \cite{Mjd1,Saffbk,Burg1}) are worked
out. Then in \S4.3 we see how these are applied to the
$(\alpha,\bchi)$ equations (\ref{DalphaDtlambda}) and (\ref{Dchi2Dt})
for the Navier-Stokes equations and find that they do indeed
correspond to fixed point solutions. In the paper by Majda \cite{Mjd1}
a series of simple examples is given which includes pure rotation, a
swirling drain and the two Burgers solutions (see \S 1C of reference
\cite{Mjd1}).  The two expressions for the vorticity $\bw ({\bf x},t)$
for the latter pair of examples are based on an $N$-dimensional heat
kernel but for the sake of simplicity in the following two subsections
we use the time asymptotic limit of this.

\subsection{Burgers vortices}

Consider a strain field $\bu =\left(-\frac{\gamma x}{2},-\frac{\gamma
y}{2}, \gamma z\right)^{T}$ with a superimposed two dimensional
velocity field $(-yf(r), xf(r),0)^{T}$ where the function $f(r)$ will
be fixed later on. The variable $r$ is given by $r^2 = x^2 +
y^2$. Then the full velocity field is
\bel{Svel1}
\bu = \left(-\frac{\gamma x}{2} -yf(r),
-\frac{\gamma y}{2}+xf(r), \gamma z\right)^{T}
\ee
and it is easily seen that the vorticity is 
\bel{Bomega1}
\bw = \left(0, 0, \omega_{3}\right)^{T},
\ee
where $\omega_{3} =2f+rf'$. Moreover the strain matrix $S$ is given by 
\bel{BS1}
S = \left(
\begin{array}{ccc}
-\frac{\gamma}{2} -\frac{xyf'}{r} & \frac{(x^{2} - y^{2})f'}{2r} & 0\\
\frac{(x^{2} - y^{2})f'}{2r} & -\frac{\gamma}{2} +\frac{xyf'}{r} & 0\\
0 & 0 & \gamma
\end{array}
\right)
\ee
Immediately we see that ${\bf e}_{3} = (0,0,1)^{T}$ 
and $\lambda_{3} = \gamma$ and the other two eigenvalues can easily 
be computed  
\bel{Sevs1}
\lambda_{1,2} = \frac{-\gamma \pm rf'}{2}
\ee
with their corresponding eigenvectors lying in the horizontal plane.
Now for the Euler equations one can simply take $f = \frac{1}{r^2}
\int_{0}^{r}s\omega_{3}(s)\,ds$ and so $\bw = \left(0, 0, \omega_{3}
(r)\right)^{T}$. For instance, there is a solution of the
three-dimensional Euler equations on compact support whose amplitude
increases exponentially while its support decreases exponentially
\cite{Mjd2}.

For the Navier-Stokes equations, however, there exists a form of
$f(r)$ in which dissipation and stretching balance which, in the limit
$t \rightarrow \infty$, is given by \cite{Mjd1}
\bel{Sf1}
f(r) = \frac{1-\exp \left(-a r^2\right)}{r^2}u_{0}.
\ee
where $a = \gamma/4\nu$. This is a profile which has its
maximum when $r = 0$.  Note that when $r = 0$ then $f(0)= a$,
$\omega_{3}(0) = 2a$ whereas $\left[rf'(r)\right]_{r=0}= 0$. 
Near $r = 0$, $rf' \sim -2a^{2}r^{2} < 0$ and so we have two negative
eigenvalues and one positive for this form of solution. Also the total
strain at $r = 0$ is
\bel{Stotal1}
\sum_{i,j}S_{ij}^{2} 
= \frac{1}{2}\left(3\gamma^{2} + (rf')^2\right)_{r=0} =
\frac{3\gamma^{2}}{2}.
\ee
Hence if $\nu \ll 1$ then $a \gg \gamma$ and so the vorticity
$\omega_{3}(0) = 2a$ is much larger than the total strain. An $x-y$
plane contrast cross-section of the vorticity field for a Burgers
vortex is produced in figure 1a.

\subsection{The Burgers shear layer solution}

Consider a jet $\bu = (0, -\gamma y, \gamma z)^{T}$ which compresses
in the $y$-direction but expands in the $z$-direction. Now impose a
velocity field $v(y)$ on the $x$-direction so that
\bel{Svel2}
\bu = \left(v(y), -\gamma y, \gamma z\right)^{T}.
\ee
\bel{Bomega2}
\bw = \left(0, 0, \omega_{3}\right)^{T},
\ee
where $\omega_{3} = -v'(y)$. For the Navier-Stokes equations, in the
limit $t \rightarrow \infty$, if $v(y)$ is taken to be 
\bel{vdef}
v(y) = \frac{1}{\sqrt{2\pi}}
\int_{-\infty}^{y\sqrt{2a}}v(0)\exp (-s^2)\,ds
\ee
with $a = \gamma/4\nu$ then 
\bel{BVSomega}
\omega_{3} = \sqrt{\frac{2a}{\pi}}
\exp(-2a y^{2})\int \omega_{0}(s)\,ds.
\ee
This the {\em Burgers shear layer} for which 
the strain matrix $S$ is given by 
\bel{BS2}
S = \left(
\begin{array}{ccc}
0 & \frac{v'}{2} & 0\\
\frac{v'}{2}& -\gamma & 0\\
0 & 0 & \gamma
\end{array}
\right).
\ee
As before we see that ${\bf e}_{3} = (0,0,1)^{T}$ 
and $\lambda_{3} = \gamma$ and the other two eigenvalues are  
\bel{Sevs2}
\lambda_{1,2} = \frac{-\gamma \pm \left(\gamma^{2} 
+ v'^{2}\right)^{1/2}}{2}
\ee
with their corresponding eigenvectors lying in the horizontal plane. 
In this case the total strain is 
\bel{Stotal2}
\sum_{i,j}S_{ij}^{2}  = 2\gamma^{2} + \frac{v'^{2}}{2}.
\ee
When $\nu \ll 1$ then $a \gg \gamma$ and therefore $|v'|^{2} \gg
\gamma^{2}$. Hence when the vorticity is high then so is the total
strain, in contrast with the vortex tube in the previous subsection.
An $x-y$ plane contrast cross-section of the vorticity field for a
Burgers stretched shear layer is produced in figure 1b,

\subsection{The $(\alpha,\chi)$ equations for both tube and shear layer}

Taking the third component of the velocity version of the
Navier-Stokes equations it is easy to show in both the above cases 
\bel{axialp} 
\frac{\partial p}{\partial z} = -\gamma z 
\ee 
and hence
\bel{alphab2} \alpha_{p} = -\gamma^{2},\hspace{2cm}\bchi_{p} = 0.  
\ee
In addition, using the solution for $\bw$ and the form of $S$, in both
cases 
\bel{alphab1} \alpha = \gamma,\hspace{2cm}\bchi = 0, 
\ee 
with corresponding angle 
\bel{{alphab3}} \phi = 0, 
\ee 
reflecting the fact that there is exact alignment between $\bw$ and
${\bf e}_{3}$ with this fixed point sitting on the $\alpha$-axis.
Clearly, in terms of the angle {\em we can make no distinction between
the tube and the shear layer}, nor is there any distinction between
the cases of alignment or anti-alignment of $\bw$ and an eigenvector
of $S$. Moreover, the unit vectors $\bxi$ and $\bs$ are given by $\bxi
= {\bf\hat{k}}$ and $\bs =0$ so $\nabla\bxi = 0$. This means that when
$\bmu = 0$ and $\lambda = \gamma^{2}$, the relevant $(\alpha,\bchi)$
equations given in (\ref{DalphaDtlambda}) and (\ref{Dchi2Dt}) are
satisfied by equations (\ref{alphab2}) and (\ref{alphab1}) for which
\bel{Lfp} 
\frac{D\alpha}{Dt} = 0,\hspace{2cm}\frac{D\bchi}{Dt} = 0.
\ee 
The problem of stability is considered in the next section; there we
will show that this fixed point is indeed stable provided $\lambda$
and $\bmu$ remain constant.  Given the fact that both Burgers cases
are essentially 2D velocity fields superimposed on three-dimensional
strain fields, any example of this type will produce a strain matrix
$S$ with a similar block diagonal form as in (\ref{BS1}) and
(\ref{BS2}) with a corresponding eigenvector ${\bf e}_{3}$ which is
parallel to the vorticity vector $\bw$. All examples of this type
therefore have $\phi = 0$.  Moffatt, Kida and Ohkitani \cite{MKO} have
used asymptotics on the Burgers vortex where they allow a certain
small asymmetry from the axisymmetric solution (\ref{Sf1}) but they
point out that only the exact symmetric solution is known.

What is not proved, however, is that if the system is attracted to a
stable Lagrangian fixed point solution with an orientation near $\phi
= 0$ then this must automatically correspond to either of the
Burgers-like solutions sketched above; there may be other unknown
structures which fall into this category (see the comments at the end
of \S2 on Hill's spherical vortex).  The lack of certainty in the
topology reflects the problem of not having enough dynamic angles in
the system.

\section{A mechanism for the formation of thin structures in
Navier-Stokes flows}

So far we have made no assumptions. Let us now discuss some
approximations and put forward the following theoretical picture which
is consistent with the formation of thin structures in isotropic
Navier-Stokes turbulence.  Over all calculations of this type lies the
heavy shadow of the technical question of regularity, which is still 
an open problem. The fact that alignment in mature turbulent flows is
being discussed, however, means that a sufficient degree of regularity
is being imputed to the solutions anyway.  From now on we assume that
the flow is regular and that all the necessary quantities are bounded.
>From \S4 we know that when the flow assumes a Burgers structure,
$\lambda$ and $\bmu$ take the values
\bel{thin}
\lambda = \gamma^{2},\hspace{2cm}\bmu = 0.
\ee
In the following, however, we do {\em not} assume that $\lambda$ and
$\bmu$ take the values given in (\ref{thin}) as this is tantamount to
assuming the answer.  Nevertheless, the highly simplified form that
$\lambda$ and $\bmu$ take when the flow assumes a Burgers state
motivates us to {\em assume that the variables $\alpha,~\bmu,~\lambda$
and $\bmu$ have reached a simultaneous equilibrium in some connected
region of vorticity whose growth has been controlled by
dissipation}.  First we want to see if the `fixed point' values
taken by $\alpha$ and $\bchi$ at this equilibrium correspond to a
small angle $\phi_{0}$ and secondly whether this fixed point is
stable. To investigate this stability question we begin by assuming
that $\lambda$ and $\bmu$ are constant in the four equations in
(\ref{DalphaDtlambda}) and (\ref{Dchi2Dt})
\bel{zetaev3}
\frac{D\alpha}{Dt} = \chi^2 - \alpha^2 + \nu \Delta\alpha + \lambda ,
\ee
\bel{equil1}
\frac{D\bchi}{Dt} = -2\alpha \bchi + \nu \Delta\bchi + \bmu 
\ee
where, for reasons of simplicity, we have written these without the
quantity $|\nabla\bxi|^{2}$.  These terms are the least important and
are dealt with in the Appendix later.  Fixed points in 
$(\alpha_{0},\bchi_{0})$ occur at
\bel{fp1}
2\alpha_{0}^2 = \lambda +\sqrt{\mu^2 + \lambda^2},\hspace{2cm}
\chi_{i,0} = \frac{\mu_{i}}{2\alpha_{0}},
\ee
where $\mu =|\bmu|$. Hence there are two fixed points in 4-space,
corresponding to the two roots of $\alpha_{0}$ in (\ref{fp1}). Without
the Laplacian term it is easy to show that the eigenvalue stability
problem is
\bel{ev1}
\left(\Lambda + 2\alpha_{0}\right)^{2}
\left[\left(\Lambda + 2\alpha_{0}\right)^{2} + 4\chi_{0}^{2}\right] 
= 0, 
\ee
thereby giving the four roots
\bel{ev2}
\Lambda = -2\alpha_{0}~~\mbox{(twice)},\hspace{2cm}
\Lambda = -2\left(\alpha_{0}\pm i\chi_{0}\right).
\ee
These roots correspond to an unstable fixed point for the negative
root for $\alpha_{0}$ and a stable one for the positive root.  In the
latter case, there is exponential contraction in two of the directions
in the 4-space with a stable spiral in the other two directions.  When
the Laplacian is included we look at the stability of linearized
solutions of the type $\exp\left(i{\bf k}\cdot{\bf x} + \Lambda
t\right)$ around $(\alpha_{0},\bchi_{0})$. The only difference
this makes to (\ref{ev2}) is that $2\alpha_{0} \rightarrow 2\alpha_{0}
+\nu k^{2}$. We conclude that provided $\alpha_{0}>0$ the equilibrium
solution is stable to the disturbance of all wavenumbers. For the
half-plane the two fixed points are
\bel{fpa1}
\sqrt{2}\alpha_{0} = \pm \left[\lambda +\sqrt{\mu^2 
+ \lambda^2}\right]^{1/2}
\hspace{2cm}
\sqrt{2}\chi_{0} = \left[-\lambda +\sqrt{\mu^2 
+ \lambda^2}\right]^{1/2}.
\ee
The stable spiral is shown in figure 2 which is projection onto the
variables $\chi_{1}$ and $\alpha$.  Defining $m$ to be
\bel{mdef}
m= \frac{\mu}{|\lambda|},
\ee
the angle $\phi_{0}$ corresponding to the stable fixed point is (the
$+$ sign is for $\lambda > 0$ and the $-$ sign for $\lambda < 0$)
\bel{phidef1}
\tan \phi_{0}^{\pm} = \frac{\sqrt{m^{2}+ 1} \mp 1}{m},
\ee
which can be simplified to
\bel{phidef2}
\tan 2\phi_{0}^{\pm} = \pm m.
\ee
Now the exact Burgers solutions both correspond to $m = 0$ and,
because $\alpha = \gamma > 0$, they correspond to an {\em attracting}
fixed point of the system. We note however that when it takes nonzero
values, $m$ is dependent only on the {\em ratio} of $\mu$ and
$|\lambda |$. Although we have no hard information on the magnitude of
$m$, in fact the angle $\phi_{0}^{+}$ is relatively insensitive to
this magnitude. For $\lambda > 0$, if $m \approx 0$ then $\phi_{0}^{+}
\approx 0$ but if $m \approx 1$ then $\phi_{0}^{+} \approx
\pi/8$. Even if $m =\infty$ then $\phi_{0}^{+} = \pi/4$. Hence at
worst, $\phi_{0}^{+}$ lies in a $45^{o}$ cone.  Therefore, even when
$m > 0$ we are still close to alignment.  More generally, $\bmu$ and
$\lambda$ derive respectively from forming the vector and scalar 
products of the same set of functions so, in an isolated region of
vorticity $\Omega$, we might expect that on the spatial average over a
random set of points in $\Omega$, $m$ would take the value $m \approx
1$.  It is also possible that because $\bmu$ is formed from a cross
product while $\lambda$ is formed from a dot product then at the
natural angle of alignment the $\mu$ term would be the weaker of
the two making $m < 1$. We therefore conclude that over the spatial
average within a small intense region for $\lambda > 0$
\bel{fp5}
\phi_{0}^{+} \approx \pi/8
\ee
but that $\phi_{0}^{+}$ may indeed be somewhat smaller than this.
Experiments \cite{ATS1,ATS2,ATS3,ATS4} and simulations
\cite{AKKG,She1,VM1,VM2,RM,BMC} generally measure the cosine of the
angle $\phi$, often observing that a bunching around $\cos\phi \approx
1$ in their PDF's is a demonstration of alignment. For 
$\phi_{0}^{+} \approx \pi/8$ we have 
\bel{pdf1}
\cos \phi_{0}^{+} \approx 0.92.
\ee
This value of $m$ therefore produces fairly close alignment. This is
consistent with the region finding an equilibrium shape near to a thin
Burgers-like structure (for which the exact value of $m$ is zero)
which corresponds to $\lambda > 0$. When $\lambda < 0$, however,
\bel{phiminus1} \phi_{0}^{-} = \pi/2 - \phi_{0}^{+} \ee and the
equivalent value of $\phi_{0}^{-}$ is $3\pi/8$. Hence vortex lines are
badly misaligned in this case.

\section{Conclusion}

The thin vortical structures which are observed in numerical
simulations of turbulence have been one of the most intriguing visual
manifestations of the complexity for which turbulence, for better or
for worse, has become a byword in the last two decades.  Our results
are consistent with near Burgers-like structures (see figure 1)
forming out of those connected parts of the flow where $\lambda$ and
$\bmu$ equilibrate and where $\lambda >0$.  The fact that the stable
equilibrium occurs in a region of $\alpha_{0} > 0$ means that there is
a preference for vortex stretching \cite{GIT1}.  To say that these
regions consist of precisely tubes and/or sheets is to oversimply the
matter; more generally the individual structures probably have a
fractal dimension which lies somewhere between one and two
corresponding to tubes and between two and three corresponding to
sheets \cite{Chorin1,Frischbk,beta,JDGfrac}.  Nevertheless, tubes and
sheets as extreme limits of the topology suffice as an approximate
description despite the fact that their local interactions and
constant metamorphosis produces levels of geometric complexity which
are beyond our understanding at the moment.  This raises several
questions regarding the geometric consequences of vorticity alignment:

\begin{enumerate}

\item Can all solutions for which $\phi \approx 0$, which are also
Lagrangian fixed points of the $(\alpha,\bchi)$ equations, be
categorized as `thin' in the sense that they are quasi 1D or 2D?
While we have characterized these structures by the angle $\phi_{0}
({\bf x},t)$ which lies between $\bw$ and $S\bw$, are there other
dynamic angles which characterize the topology more specifically?

\item Are the Burgers solutions or their generalizations unique among
the set of thin solutions?

\item What is the behaviour of $m = \mu/|\lambda|$ defined in
(\ref{mdef}) and what determines the sign of $\lambda$?  The
assumption that $\lambda$ and $\bmu$ are roughly constant in some
regions is based on the assumption that (consistent with the fact that
they behave this way for the two Burgers solutions) a balance occurs
at the deepest scales between the pressure and the viscous terms
expressed through $P$, $S$, $\nabla S$ and $\nabla\bw$. Such an
assumption would be hard to prove directly while no regularity proof
exists.

\item How are $\alpha$ and $\bchi$ affected when there are local
interactions; for instance, when two sheets interact and wrap up to
become tubes \cite{VM2}?  It is possible that large values of
$\lambda$ correspond more to sheets than tubes since the former have
higher strain.  The experiments of Andreotti {\em et al} \cite{Cd2}
creating Burgers vortices suggests that the tubes formed from the
roll-up of sheets have a long lifetime compared to the roll-up
time. This suggests that two time scales may be involved; the first
being the time it takes for the orbit in $(\alpha,\bchi)$ space to
stabilize and the second the time over which $\lambda$ and $\bmu$
remain constant before themselves decaying due to dissipative effects.

\item What role does the Hessian matrix $P$ play in these processes?
Is there a consistent pattern in the sign of $\alpha_{p}$?

\end{enumerate}
As far as question 1 is concerned, it is clear that one angle is by no
means sufficient to adequately describe the topology. There are two
other apparent natural angles in the system.  The first of these is
based on the matrix $\omega_{i,j}$ instead of $S_{ij}$ which we
discuss only in the Euler case.  Define
\bel{Bequ1}
\aw = \frac{\omega_{i}\omega_{i,j}\omega_{j}}{\bw\cdot\bw}  = 
\frac{\bw\cdot (\bw\cdot\nabla)\bw}{\bw\cdot\bw} 
\ee
and then use the general result (\ref{Aequn2}) to show that 
\bel{Bequ2}
\frac{~D\aw}{Dt} = \bw\cdot\nabla\alpha .
\ee
(\ref{Bequ2}) shows that points moving with the flow in regions where
$\alpha = \mbox{const}$ (or even just spatially independent) also have
$\aw$ as a constant of the motion. Now define 
\bel{Bequn3}
\bchiw = \frac{\bw\times (\bw\cdot\nabla)\bw}{\bw\cdot\bw} 
\ee
then 
\bel{Bequn4}
\frac{~D\bchiw}{Dt} = (\bw\cdot\nabla)\bchi+2\aw\bchi -2\alpha\bchiw
+\frac{2\bsig\times(\bw\cdot\nabla)\bw}{\bw\cdot\bw}.
\ee
When $\phi = 0$ then $\bsig = S\bw = \alpha \bw$ and $\bchi = 0$ and so
\bel{Bequn5}
\frac{~D\bchiw}{Dt} = 0.
\ee
In consequence the angle $\psi$ between $\bw$ and $(\bw\cdot\nabla)\bw$
for points travelling with the flow is given by 
\bel{Bequn6}
\tan \psi = \frac{\mbox{const}}{\aw}.
\ee
$\psi$ therefore decreases in regions where $\aw$ is stretching.  The
second angle concerns the Hessian matrix of the pressure $P$
\bel{con2}
\tan\phi_{P} = \frac{|\bw\times P\bw|}{\bw\cdot P\bw}, 
\ee 
but we have no separate knowledge of the evolution of $P$. In fact $P$
and $S$ are less independent than one would think.  A consequence of
equation (\ref{conv2}) is that when $\bw$ aligns with an eigenvector
of $S$ then $\bw$ also aligns simultaneously with an eigenvector of
$P$. 

The lack of knowledge of $m$ raised in question 3 above is balanced in
part by the insensitivity of $\phi_{0}$, the attracting angle, to the
value of $m$ through the relation ($\lambda > 0$) 
\bel{con1}
\tan2\phi_{0}^{+}= m.  
\ee 
There is no reason, a priori, why exact alignment at $m= 0$ should be
attracting. Nevertheless, values of $m >0$ produce attracting
orientations which are still close enough to alignment to suggest that
we are near a thin structure which we conjecture is perhaps twisted in
some way. As we showed in the last section, different regions may take
different spatial values of $m$ but their associated angles may still
be close enough to look the same in observations or simulations. $m$
obviously contains geometrical information about the local and
relative orientation of vortex lines in the sense of the Frenet-Serret
equations \cite{CPS,GPS} but these need information on $\nabla \bw$
which we do not have.  It is possible that more can be said about $m$
using a scaling argument. 

What the formulation of this paper does not do is differentiate
between tubes and sheets; the alignment process is the same for both
and we have no other information which distinguishes them. In fact,
one very important set of processes, not accounted for by the dynamics
of $\phi$ alone, are the interaction processes between one sheet and
another.  The Kelvin-Helmholz instability, which is an Euler
phenomenon, is a well known mechanism through which it is thought that
two dimensional sheet-like objects wrap up to become tubes
\cite{Kerr3,VM2,Sulem1,RM,KOK}. Such local interaction processes are
extremely subtle and if they are to be accounted for dynamically then
certainly more than one angle would be needed.

The thin structures observed in simulations take up no more than a few
percent of the total flow volume indicating that vorticity is not
distributed evenly. This may be one of the reasons behind the failure
to prove regularity. The methods used have been based upon attempts to
control the {\em global} enstrophy, $\int_{\Omega}|\bw|^{2}\,dV$,
which have foundered on the problem of the dissipation being too weak
to control the estimated nonlinear terms (see references in
\cite{JDGbk}).  These latter terms have their origin in the vortex
stretching term, $\int\omega_{i}S_{ij}\omega_{j}\,dV$, and have to be
estimated by standard Sobolev inequalities. The failure to control the
global enstrophy for more than a finite time by this method probably
has its origins in the use of volume integration in which
$L^{2}$-norms average over, and perhaps miss, the spatial structures
where the nonlinearity has been strongly depleted by
alignment. Constantin's Biot-Savart integral formula for $\alpha$ in
terms of a triad of vectors related to $\bxi$ illustrates how
nonlinearity is depleted inside the volume integral when significant
alignment takes place \cite{Constantinreview}.

The general picture that emerges from this analysis is that the
variables $(\alpha,\bchi)$ seem to form a natural pair of variables in
which to express the dynamics of vortex formations.  In both the Euler
and Navier-Stokes cases negative values of $\alpha$ are repelling and,
without interference from the pressure, $\phi$ would limit to
zero. Any initial state corresponding to a negative value of $\alpha$
would require the system to go through the $\phi = \pi/2$ stage before
becoming a Burgers-like shear layer or tube if indeed it is driven
that far. The difference between the two is that for the Euler
equations there is no stable equilibrium point provided by the
viscosity and the whole process could reverse, although Ohkitani and
Kishiba \cite{Ohk2} report that they observe an alignment between
$\bw$ and the third eigenvector of $P$. A systematic study of
simulation data for the three-dimensional Euler equations is underway
to understand the properties of the more general behaviour of the
Hessian matrix $P$.

One curious conclusion from (\ref{alphaev2}) in the Euler equation
case, for a Lagrangian particle element initially at ${\bf X} = {\bf
x}(0)$, is that if $\alpha ({\bf X},0)<0$ and $\phi({\bf X},t) <
\pi/4$ with $\alpha_{p} > 0$ then $\alpha \rightarrow -\infty$ in a
finite time.  In contrast, if $\alpha ({\bf X},0) > 0$ and $\phi({\bf
X},t) > \pi/4$ with $\alpha_{p} < 0$ then $\alpha \rightarrow \infty$
in a finite time. This is consistent with the results of Ng and
Battercharjee \cite{NgBat} who have studied the Euler equations under
high symmetry conditions which makes the problem quasi
one-dimensional.

Past simulations have shown that initially $\bw$ aligns with the
largest eigenvector of $S$ but as the turbulence becomes more mature
$\bw$ aligns with the second eigenvector \cite{AKKG,She1,VM2} as a
global average. Tsinober, Shtilman and Vaisburd \cite{ATS4} have
recently reported, however, that {\em locally} they see significant
alignments between $\bw$ and {\em both} the first and the second
eigenvectors of $S$.  They also report that the vorticity does not
have to be too intense and they produce evidence to suggest that the
background field is not Gaussian.

\par\bigskip\noindent {\bf Acknowledgments:} We are grateful to
Charles Doering, Robert Kerr, Andrew Majda, Itamar Procaccia, Len
Sander, Edriss Titi and Arkady Tsinober for several helpful
suggestions. JDG warmly thanks the trustees of the Sir Siegmund
Warburg Foundation in London for their financial support during his
nine month stay at the Weizmann Institute of Science.  Matthew
Heritage is thankful to the UK EPSRC Mathematics Committee for the
award of a studentship.

\appendix
\section{The effect of the misalignment terms} 

For the $|\nabla\bxi|^{2}$ terms left out of the discussion in \S5,
it would be preferable if we had five evolution equations in $\alpha$,
$\chi_{i}$ and $\nabla\bxi$ and not just four but we have been unable
to discover any delicate cancellations in the material derivative of
the last variable. Instead, we include this as a constant term in the
equilibrium point analysis above.  For a vortex line, $\nabla\bxi$ is
called the misalignment and plays an interesting role. Constantin,
Procaccia and Segel \cite{CPS} note that
\bel{misal2}
|\nabla\bxi|^{2} = |(\bxi\cdot\nabla)\bxi|^{2}
+ |({\bf \hat{n}}\cdot\nabla)\bxi|^{2}
+ |({\bf \hat{b}}\cdot\nabla)\bxi|^{2}
\ee
where the first term on the RHS of (\ref{misal2}) is the square of the
curvature and the second and third are the squares of the the lack of
parallelity between a vortex line and neighbouring lines in the normal
$({\bf \hat{n}})$ and binormal $({\bf \hat{b}})$ directions
respectively. Galanti, Procaccia and Segel \cite{GPS} have produced
theoretical and numerical arguments which indicate that the stronger
the vorticity and the greater the curvature of a vortex line the
stronger the stretching and therefore the more liable it is to
straighten. To show how this is the case we define 
\bel{adef}
a = \nu |\nabla\bxi|^{2}\hspace{1cm}
\at = \alpha - a\hspace{1cm}
\lt = \lambda + a^{2}.
\ee
The $(\alpha,\bchi)$ equations become
\bel{neweqn1}
\frac{D\at}{Dt} = \chi^2 - \at^2 + \lt ,
\ee
\bel{neweqn2}
\frac{D\bchi}{Dt} = -2\at\bchi + \bmu .
\ee
The stability problem is unchanged and fixed points come from 
\bel{FPmis1}
\at_{0}^{2} = \frac{\lt + \left(\lt^{2}+ \mu^{2}\right)^{1/2}}{2}.
\ee
>From (\ref{FPmis1}), $\at_{0}^{2} \geq \lt \geq a^{2}$ when $\lambda >
0$ and so $\alpha_{0} \geq 2a$ for positive values of $\alpha_{0}$.
Moreover, when considering the angle of orientation of the equilibrium
point $\tilde{\phi}_{0}$ we see that
\bel{newang1}
\tan 2\tilde{\phi}_{0} 
= \frac{|\mu|}{|\lt |} 
= \frac{|\mu|}{|\lambda +a^2 |} \leq m.
\ee
Hence
\bel{newang2}
\tan 2\tilde{\phi}_{0} \leq \tan 2\phi_{0}.
\ee
Equation (\ref{newang2}) shows that nonzero values of
$|\nabla\bxi|^{2}$ act to make the angle smaller than when it is
excluded. Of course the angle $\phi$ and $|\nabla\bxi|^{2}$ are
different ways of expressing the same effect. We actually need another
differential equation to form an accurate picture and this, in turn,
needs information on $\nabla S$ and $\nabla\bw$.  The above
calculation shows that for the $(\alpha,\bchi)$ equations alone large
values of $|\nabla\bxi|^{2}$ force stronger alignment in $\phi$,
thereby confirming the effect seen in \cite{GPS}.


\bibliographystyle{unsrt}

\newpage

\bc
{\Large\bf FIGURE CAPTIONS}
\ec

\par\bigskip\noindent
{\bf Figure 1a:} $x-y$ cross section of a Burgers vortex with the
$z$-co-ordinate pointing out of the paper in which direction points
the vorticity vector $\bw = (0,0,\omega_{3})^{T}$.  In the black and
white contrast, the whiter the area the higher the vorticity.

\par\bigskip\noindent 
{\bf Figure 1b:} $x-y$ cross section of a Burgers stretched shear
layer. As in Figure 1a, the $z$-co-ordinate points out of the paper in
which direction points the vorticity vector $\bw =
(0,0,\omega_{3})^{T}$.  In the black and white contrast, the whiter
the area the higher the vorticity.

\par\bigskip\noindent 
{\bf Figure 2:} Orbits in the $\alpha-\bchi$ phase space projected
onto the $\alpha-\chi_{1}$ plane. The spiral structure at the two
fixed points is just evident.

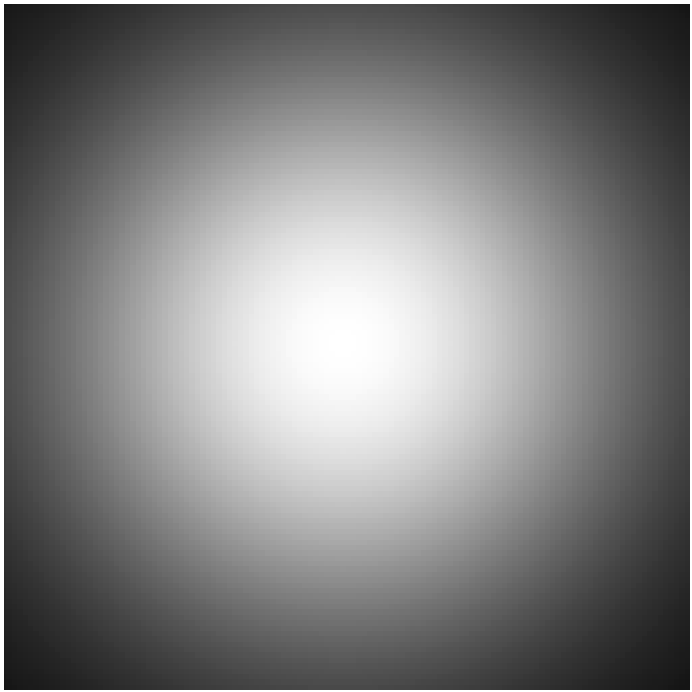
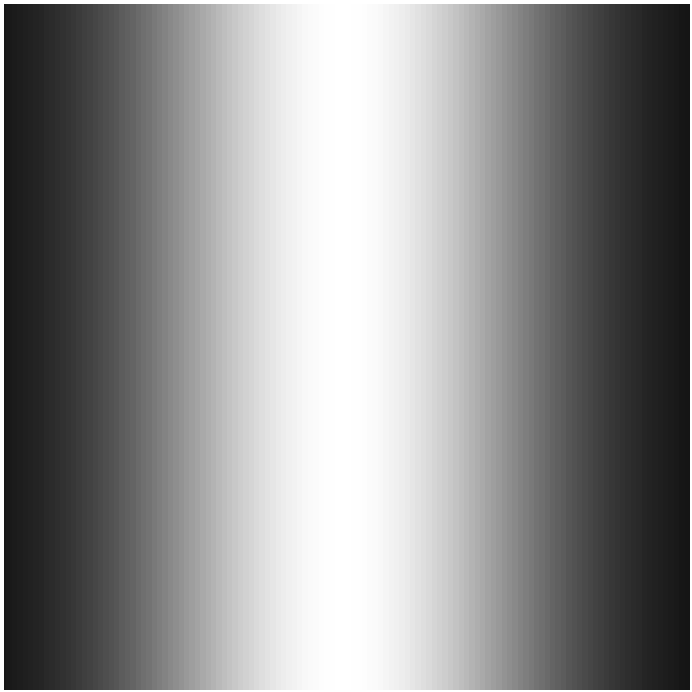
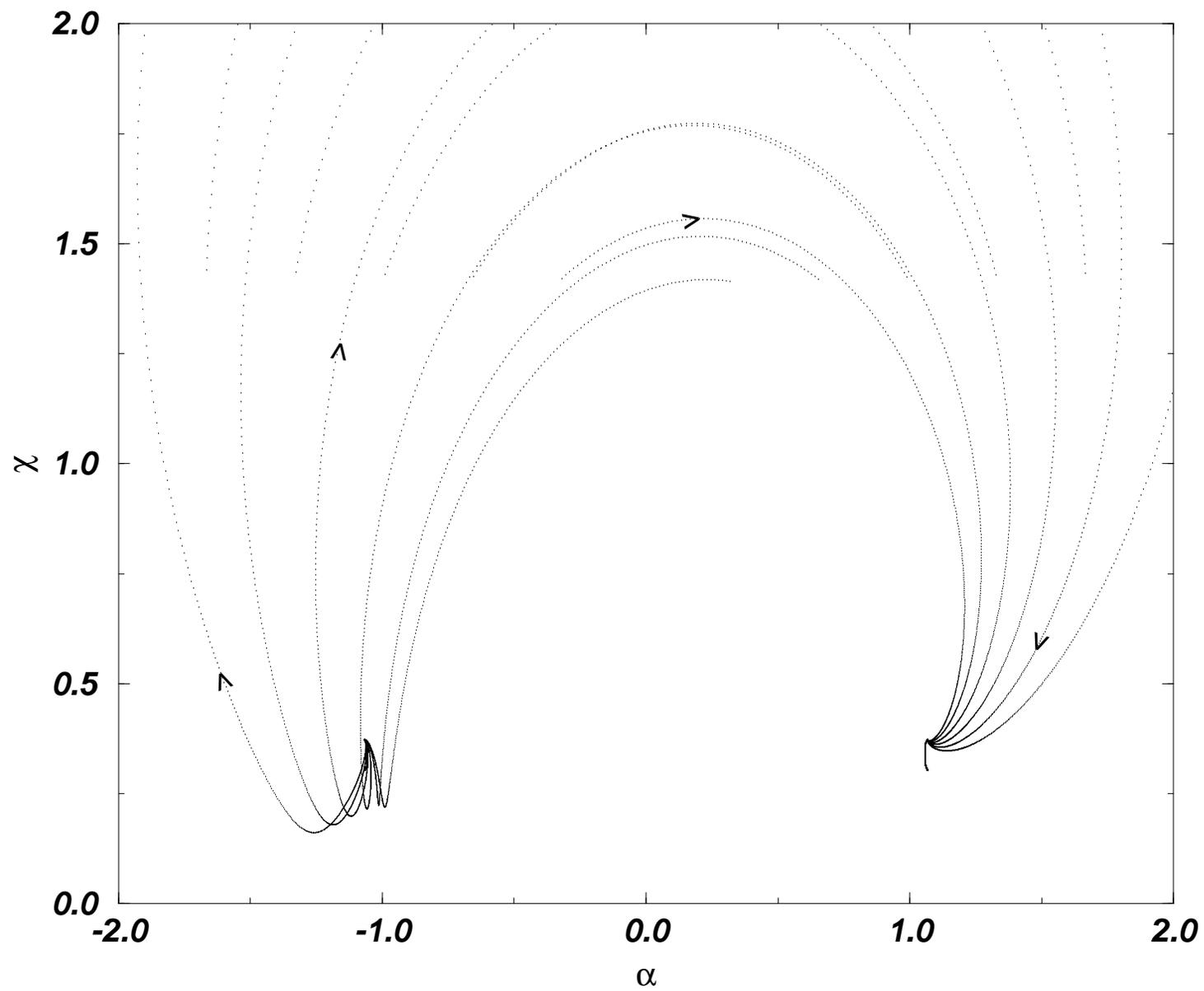
\end{document}